\documentclass[12pt]{article}
\usepackage[english]{babel}
\usepackage{amsmath,amsthm}
\usepackage{amsfonts}

\usepackage{color}

\theoremstyle{definition}

\theoremstyle{remark}

\numberwithin{equation}{section}

\begin{document}

\title{\textbf{Some aspects of the $\boldsymbol m$-adic analysis\\ and its applications\\ to
$\boldsymbol m$-adic stochastic processes}}

\author{M.V.\,Dolgopolov, %
\thanks{Samara State University, e-mail: mikhaildolgopolov@rambler.ru%
} \\
and \\
 A.P.\,Zubarev%
\thanks{Samara State University, e-mail: apzubarev@mail.ru%
}}
\maketitle
\begin{abstract}

In this paper we consider a generalization of analysis on $p$-adic numbers
field to the $m$ case of $m$-adic numbers ring. The basic statements, theorems and formulas
of $p$-adic analysis  can be used for the case
of $m$-adic analysis without changing. We discuss basic
properties of $m$-adic numbers and consider some properties of $m$-adic integration and $m$-adic
Fourier analysis.
The class of infinitely
divisible $m$-adic distributions and the class of $m$-adic stochastic
Levi processes were introduced. The special class of $m$-adic
CTRW process and  fractional-time $m$-adic random walk as the diffusive limit of it is considered. We
 found the asymptotic behavior of the probability measure of initial
distribution support for fractional-time $m$-adic random walk.
\end{abstract}

\section{Introduction}

Up to recent time, ultrametric spaces were known only in mathematical
literature. A classic example of such space is the $p$-adic numbers
\cite{key-1,key-2} field. During the last three decades the interest
in ultrametric models research has risen in various fields of physics,
biology, economics and sociology, such as models of spin glass, biopolymers,
fractal structures as well as in optimization theory, taxonomy, evolutional
biology, cluster and factor analysis etc. (refer to the surveys \cite{key-3}
and \cite{key}). In recent years the scientific interest laid also
in the field of the dynamic models for the ultrametric structures,
in particular the models of ultrametric diffusion (see \cite{key-4,key-5,key-6,key-7,key-8}
and references therein), which are directly related to the description
of the conformational rearrangements of the proteins dynamics as well,
apparently may be related to the random processes modeling in the
complicated biological and socio-economic systems \cite{key-9,key-10,key-11}.

The ultrametricity of space is closely related to a concept of hierarchic
structure over space \cite{key-3}. For complex systems modeling hierarchic
structures may appear both on the space of states of the system and
on the space of objects constituting the system. E.\,g., one describes
dynamics of conformation rearrangements of biopolymer macromolecules,
multiple local minima of free energy can be regarded as highly degenerated
space of the system states possessing the hierarchic structure \cite{key-4,key-5,key-6,key-7}.

An adequate mathematical tools to formalize description of mathematical
modeling of the complex systems hierarchic structures are required.
One such tool is a $p$-adic analysis developed during the last three
decades by the group of Academician V.S.\,Vladimirov in Steklov Mathematical
Institute. The main idea of application of $p$-adic analysis at mathematical
modeling of complex hierarchically organized systems is that the field
of $p$-adic numbers $\mathbf{Q}_{p}$ has a natural regular indexed
hierarchic structure with the degree of regularity equal to certain
prime number $p$ and thus a configuration space of such systems can
be adequately described by $p$-adic coordinate space.

Nevertheless, it is worth to mention that, strictly speaking, the
mathematical tools of $p$-adic analysis is directly applicable for
describing a hierarchically organized system only in the case when
the system possesses regular indexed hierarchic structure and degree
of regularity of the hierarchic structure is a prime number $p$.
The analysis on general ultrametric spaces described by oriented trees
has been developed in \cite{key-12}. Nevertheless, for specific computations
it is necessary to use explicit parametrization of the points of an
ultrametric space, in which the computational results are represented.
For example, in the case of $\mathbf{Q}_{p}$ such parametrization
has the form of $a_{n}p^{n}+a_{n-1}p^{n-1}+...$. The using of such
a parametrization for ultrametric space corresponding to an oriented
tree of general type is rather difficult.

In this paper, which is partly methodological by nature, we consider
a generalization an analysis on a field $\mathbf{Q}_{p}$ for more
general case of ring of $m$-adic numbers $\mathbf{Q}_{m}$.
Some aspects of $m$-adic analysis are considered in the
book \cite{key-14}.
The ring $\mathbf{Q}_{m}$ as well as the field $\mathbf{Q}_{p}$ is local-compact
Abelian group. Thus~for $\mathbf{Q}_{m}$ there exists a unique (up
to a normalization) translation-invariant Haar measure and dual group
of characters which allows to develop Fourier analysis and consider
the classes of pseudo-differential operators. We show that the basic
statements, theorems and formulas of $p$-adic analysis \cite{key-2,key-13}
can be used for the case of $m$-adic analysis without changing. In
section 2 we discuss basic properties of $m$-adic numbers $\mathbf{Q}_{m}$.
In sections 3 and 4 we consider some properties of $m$-adic integration
and $m$-adic Fourier analysis. In section 5 we introduce a class
of infinitely divisible $m$-adic distributions and a class of $m$-adic
random Levi processes. In section 6 we consider the special class
of random processes of uncoupled continuous time random walk type
on $\mathbf{Q}_{m}$ ($m$-adic CTRW). In section 7 we discuss the
diffusive limit of $m$-adic CTRW together with the equation of fraction-time
$m$-adic random walk. In~section~8 we find the asymptotic behavior
of probability measure of initial distribution support for a random
process described by fraction-time equation of $m$-adic random walk.

\section{The ring of $\boldsymbol{m}$-adic numbers $\mathbf{Q}_{\boldsymbol{m}}$}

Let us define a ring $\mathbf{Q}_{m}$ as the set of infinite series
\[
x=\sum_{i=r}^{\infty}a_{i}m^{i},\]
 where $r$ is an integer number, $a_{r}=1,2,3,...,m-1$, $a_{i}=0,1,2,3,...,m-1$
($i\ne r$). The set $\mathbf{Q}_{m}$ has a structure of commutative
associative ring, i.\,e. the sum and product operations are defined
in it. Let us call a sum of two $m$-adic numbers $x=\sum_{i=r}^{\infty}a_{i}m^{i}$
and $y=\sum_{i=s}^{\infty}b_{i}m^{i}$ such element $z=\sum_{i=l}^{\infty}c_{i}m^{i}$,
$l=\min\left\{ r,s\right\} $ that \begin{equation}
\sum_{i=r}^{\infty}a_{i}m^{i}+\sum_{i=s}^{\infty}b_{i}m^{i}=\sum_{i=l}^{\infty}c_{i}m^{i}.\label{sum}\end{equation}
 Let us call the product of two $m$-adic numbers $x=\sum_{i=r}^{\infty}a_{i}m^{i}$
and $y=\sum_{i=s}^{\infty}b_{i}m^{i}$ such an element $z=\sum_{i=l}^{\infty}c_{i}m^{i}$
that $l\ge r+s$ and \begin{equation}
\sum_{i=r}^{\infty}\sum_{j=s}^{\infty}a_{i}b_{j}m^{i}m^{j}=\sum_{i=r+s}^{\infty}c_{i}m^{i}.\label{prod}\end{equation}
 Comparing right and left parts of equations (\ref{sum}) and (\ref{prod})
by $\mathrm{mod}\: m^{n}$ for $n=1,2,3,...$ makes it possible to
find successively the coefficients $c_{i}$ for the sum $z=x+y$ and
product $z=xy$.

Division operation in $\mathbf{Q}_{m}$ is uniquely defined only in
case $m$ is a prime $p$. In~such case, the ring $\mathbf{Q}_{m}$
is a field of $p$-adic numbers $\mathbf{Q}_{p}$.

There exists $m$-adic pseudonorm $|x|_{m}=m^{-r}$ in $\mathbf{Q}_{m}$
for $x=\sum_{i=r}^{\infty}a_{i}m^{i}$
with $a_r \ne 0$
which satisfies the following
properties:
\begin{enumerate}
\item $|x|_{m}=0$ $\Leftrightarrow$ $x=0$,
\item $|xy|_{m}\le|x|_{m}|y|_{m}$,
\item $|x+y|_{m}\le\max\left\{ |x|_{m},|y|_{m}\right\} $.
\end{enumerate}
In the case $m$ is a prime $p$, non-eq. 2. transforms into eq. $|xy|_{p}=|x|_{p}|y|_{p}$,
and $m$-adic pseudonorm $|x|_{m}$ transforms into $p$-adic norm
$|x|_{p}$. Note that for any $x\in\mathbf{Q}_{m}$ and any $r\in\mathbf{Z}$,
the following equation takes place: \begin{equation}
|m^{r}x|_{m}=m^{-r}|x|_{m}.\label{prod_m}\end{equation}
 Pseudonorm $|x|_{m}$ induces metrics $d_{m}(x,y)=|x-y|_{m}$ satisfying
the following properties:
\begin{enumerate}
\item $d_{m}(x,y)=0$ then and only then if $x=y$,
\item $d_{p}(x,y)\le\max\{d_{p}(x,z),\mathrm{\;\;}d_{p}(y,z)\}$ for any
$x$, $y$ and $z$ from $\mathbf{Q}_{m}$.
\end{enumerate}
We call the subset $B_{r}(x_{0})=\left\{ x\in\mathbf{Q}_{m}:|x-x_{0}|\le m^{r}\right\} $
a disk of radius $m^{r}$ with the center at the point $x_{0}\in\mathbf{Q}_{m}$,
and subset $S_{r}(x_{0})=\left\{ x\in\mathbf{Q}_{m}:|x-x_{0}|=m^{r}\right\} $
-- a~circumference of radius $m^{r}$ with the center at the point
$x_{0}\in\mathbf{Q}_{m}$. Also we denote $B_{r}(0)\equiv B_{r}$,
$S_{r}(0)\equiv S_{r}$, $B_{0}\equiv Z_{m}$. Due to the same metric
structure of both $\mathbf{Q}_{m}$ and $\mathbf{Q}_{p}$ the following
statements for $\mathbf{Q}_{m}$ take place \cite{key-2}:
\begin{enumerate}
\item All disks and circumferences in $\mathbf{Q}_{m}$ are closed and open
sets.
\item All points of a disk is also its center.
\item If two disks have a common point, then one of them is contained in
another.
\item All sets open in $\mathbf{Q}_{m}$ are countable unions of nonintersecting
disks.
\item $\mathbf{Q}_{m}$ has zero dimension.
\item $\mathbf{Q}_{m}$ is separable totally disconnected, Hausdorf, locally
compact.
\end{enumerate}

\section{Measure and integration on $\mathbf{Q}_{\boldsymbol{m}}$}

As $\mathbf{Q}_{m}$ is a local compact commutative group there exist
an additive translation invariant measure $\mu(B\in\mathbf{Q}_{m})$
\cite{key-2}. This measure is uniquely determined by properties:
\begin{enumerate}
\item $\mu\left(Z_{m}\right)=1$ (normalization),
\item $\forall$$a\in\mathbf{Q}_{m}$, $\forall$ $B\in\mathrm{B}$ $\mu\left(B\right)=\mu\left(B+a\right)$
(translational invariance),
\item $\forall$$\left\{ B_{i}\in\mathrm{B}\right\} $, $B_{i}\bigcap B_{j}=\emptyset$,
$i\ne j$ $\Rightarrow$ $\mu\left(\bigcup_{i}B_{i}\right)=\sum_{i}\mu\left(B_{i}\right)$
(additivity).
\end{enumerate}
Note that the measure satisfies the following property $\mu\left(m^{r}A\right)=m^{-r}\mu\left(A\right)$
but in general $\mu\left(xA\right)\ne|x|_{m}\mu\left(A\right)$. For
example in the case $m=4$ we have $2\cdot B_{0}\subset B_{0}$. On
the other hand \[
2\cdot B_{0}=2\cdot\left(\left(0+B_{-1}\right)\bigcup\left(1+B_{-1}\right)\bigcup\left(2+B_{-1}\right)\bigcup\left(3+B_{-1}\right)\right)=\]
 \[
=\left(2\cdot B_{-1}\right)\bigcup\left(2+2\cdot B_{-1}\right)\bigcup\left(4+2\cdot B_{-2}\right)\bigcup\left(2+4+B_{-1}\right),\]
 therefore $3+B_{-1}\not\subset2\cdot B_{0}$ and $2\cdot B_{0}\subset B_{0}\backslash\left(3+B_{-1}\right)$.
By additivity $\mu\left(2\cdot B_{0}\right)\le\mu\left(B_{0}-\left(3+B_{-1}\right)\right)=\mu\left(B_{0}\right)\backslash\mu\left(3+B_{r-1}\right)=4^{r}-4^{r-1}$
and $\mu\left(2\cdot B_{r}\right)\ne|2|_{4}\mu\left(B_{r}\right)=4^{r}$.

For $B\subset\mathbf{Q}_{m}$ let us call \[
\mu^{\star}\left(B\right)=\inf_{B\subset\bigcup B_{i}}\sum_{i}\mu_{i}\left(B_{i}\right)\]
 an external measure, where the infimum is taken over all possible
coverings of the $B$ by countable set of non-intersecting disks $\left\{ B_{i}\right\} $.
$B$ is measurable if for any $\varepsilon>0$ it is possible to find
such elementary set (finite union of non-intersecting disks) $A$ that $\mu^{\star}\left((B\bigcup A)\backslash(B\bigcap A)\right)<\varepsilon$.
Let $\mathrm{B}$ be the class of all measurable subsets of $\mathbf{Q}_{m}$.
Then for any $B\in\mathrm{B}$ $\mu\left(B\right)\equiv\mu^{*}\left(B\right)$.
It is possible to prove (e.g. refer to \cite{key-15}) that $\mathrm{B}$
is $\sigma$-algebra and the measure $\mu\left(B\right)$ is $\sigma$-additive
on $\mathrm{B}$.

A complex valued function $f(x)$ on $B\in\mathrm{B}$ is measurable
if for any Borelean
$\mathrm{B} \subset \mathbf{C}$
the set
$f^{-1}(\mathrm{B})$
is measurable one. For
measurable functions $f(x)$ it is possible to define linear continuous
functional -- integral $\int_{B}f(x)d_{m}x$ of the function $f(x)$
over the measure $\mu(B)$.

The set of measurable on $B$ functions for which the integral $\int_{B}|f(x)|^{\lambda}d_{m}x$
at $\lambda\ge1$ is finite, will be denoted by $L^{\lambda}(B)$.
Let $f(x)\in L_{\mathrm{loc}}^{\lambda}(\mathbf{Q}_{m})$ if for any
measurable compact set $B\in\mathbf{Q}_{m}$ $f(x)\in L^{\lambda}(B)$.

A convolution $(f*g)(x)$ of two functions is \[
(f*g)(x)=\intop_{\mathbf{Q}_{m}}f(y)g(x-y)d_{m}y.\]

\textbf{Theorem 3.1. } \emph{Let the functions $f(x)$ and $g(x)$
are continuous and bounded on~$\mathbf{Q}_{m}$, $f(x)\in L^{1}(\mathbf{Q}_{m})$,
$g(x)\in L^{1}(\mathbf{Q}_{m})$. Then $(f*g)(x)$ exists for any
$x$. Moreower, $(f*g)(x)$ is continuous and bounded, $(f*g)(x)\in L^{1}(\mathbf{Q}_{m})$
and } \emph{\[
\intop_{\mathbf{Q}_{m}}(f*g)(x)d_{m}x=\intop_{\mathbf{Q}_{m}}f(x)d_{m}x\intop_{\mathbf{Q}_{m}}g(x)d_{m}x.\]
 }The proof is the same as in regular analysis (e.\,g. see \cite{key-16}).

Below some integrals, that we need later, are present.

If a function $f(x)$ depends only on $|x-a|_{m}$, then the integral
of it over disk $B_{r}(a)$ equals to \begin{equation}
\intop_{B_{r}(a)}f(|x-a|_{m})d_{m}x=\left(1-m^{-1}\right)\sum_{i=-\infty}^{r}m^{i}f(m^{i}).\label{int_f}\end{equation}

Let us define the function \begin{equation}
\chi_{m}(x)=\exp\left(2\pi i\left\{ x\right\} \right),\label{char}\end{equation}
 where $\left\{ x\right\} $ is fraction part of $x$ defined as \begin{equation}
\left\{ \sum_{i=r}^{\infty}a_{i}m^{i}\right\} =\left\{ \begin{array}{l}
\sum_{i=r}^{-1}a_{i}m^{i},\: r<0,\\
\\0,\: r\ge0.\end{array}\right.\label{{}}\end{equation}
 Let us evaluate the integral $\intop_{B_{r}}\chi_{m}(kx)d_{m}x$
where $k\in\mathbf{Q}_{m}$. For $|k|_{m}\le m^{-r}$ the~following relations $|kx|_{m}\le|k|_{m}|x|_{m}\le1$
and $\chi_{m}(kx)=1$ take place, therefore we have $\intop_{B_{r}}\chi_{m}(kx)d_{m}x=m^{r}$.
At $|k|_{m}>m^{-r}$ we shall make a substitution $x\to y=x-m^{r}$,
$B_{r}\to B_{r}(p^{r})=B_{r}$. Then \[
\intop_{B_{r}}\chi_{m}(kx)d_{m}x=\intop_{B_{r}}\chi_{m}(ky+km^{r})d_{m}y=\chi_{m}(km^{r})\intop_{B_{r}}\chi_{m}(ky)d_{m}y\]
 and we obtain \begin{equation}
\left(1-\chi_{m}(km^{r})\right)\intop_{B_{r}}\chi_{m}(ky)d_{m}y=0.\label{int_hi}\end{equation}
 As $|k|_{m}>m^{-r}$ (see (\ref{prod_m})) we have $|km^{r}|_{m}=|k|_{m}|m^{r}|_{m}=|k|_{m}m^{-r}>1$.
By (\ref{char})--(\ref{{}}) $\left(1-\chi_{m}(km^{r})\right)\ne0$
it follows from (\ref{int_hi}) that $\int_{B_{r}}\chi_{m}(ky)d_{m}y=0$.
Therefore we have \begin{equation}
\intop_{B_{r}}\chi_{m}(ky)d_{m}y=\left\{ \begin{array}{l}
m^{r},\mathrm{\:}|k|_{m}\le m^{-r},\\
\\0,\:|k|_{m}>m^{-r}.\end{array}\right.\label{int_hi_B_r}\end{equation}
 From (\ref{int_hi_B_r}) it follows that \begin{equation}
\intop_{S_{r}}\chi_{m}(ky)d_{m}y=\left\{ \begin{array}{l}
m^{r}(1-m^{-1}),\:|k|_{m}\le m^{-r},\\
\\-m^{r-1},\:|k|_{m}=m^{-r+1},\\
\\0,\: k|_{m}>m^{-r+1}.\end{array}\right.\label{int_hi_S_r}\end{equation}
 It is not difficult to find 
for any $f(|x|_{m})\in L_{\mathrm{loc}}^{1}(\mathbf{Q}_{m})$ \[
\intop_{B_{r}}f(|x|_{m})\chi_{m}(kx)d_{m}x=\]
 \begin{equation}
=\left\{ \begin{array}{l}
\left(1-m^{-1}\right)\sum_{r=-s}^{\infty}m^{-r}f\left(m^{-r}\right),\:|k|_{m}\le m^{-s},\\
\\\left(1-m^{-1}\right)\dfrac{1}{|k|_{m}}\sum_{r=0}^{\infty}m^{-r}f\left(\dfrac{m^{-r}}{|k|_{m}}\right)-\dfrac{1}{|k|_{m}}f\left(\dfrac{m}{|k|_{m}}\right),\:|k|_{m}>m^{-s}.\end{array}\right.\label{int_hi_f_B_r}\end{equation}

Note that the integrals (\ref{int_f}), (\ref{int_hi_B_r}), (\ref{int_hi_S_r})
and (\ref{int_hi_f_B_r}) coincide with corresponding integrals on
$\mathbf{Q}_{p}$ \cite{key-2} after substitution $p\to m$.

\section{Fourier analysis on $\mathbf{Q}_{\boldsymbol{m}}$}

The character of $\mathbf{Q}_{m}$ as an abelian additive group is
a complex-valued function $\chi(x)$ on $\mathbf{Q}_{m}$ with the
following properties:
\begin{enumerate}
\item $|\chi(x)|=1,$
\item $\chi(x)=1\:\Leftrightarrow\: x=0,$
\item $\forall x,y\in Q_{m}\:\chi(x+y)=\chi(x)\chi(y).$
\end{enumerate}
\textbf{Theorem 4.1. }\emph{The group of characters
on group $\mathbf{Q}_{m}$ is isomorphous to $\mathbf{Q}_{m}$, and
arbitrary element $\chi(x)$ from group of characters on group
$\mathbf{Q}_{m}$ is given by $\chi_{m}(kx)=\exp\left(2\pi i\left\{ kx\right\} \right)$,
}$k\in\mathbf{Q}_{m}$\emph{.}

\textbf{Proof.} Consider some character $\chi(x)$. Let $\int_{B_{r}}\chi(x)d_{m}x=c$.
If $c\ne0$, then for all $y\in B_{r}$ from the equation \[
\int_{B_{r}}\chi(x^{\prime})d_{m}x^{\prime}=\int_{B_{r}}\chi(x+y)d_{m}(x+y)=\chi(y)\int_{B_{r}}\chi(x)d_{m}x\]
 it follows that $\chi(y)=1$. Let $s=\max\left\{ r:\:\int_{B_{r}}\chi(x)d_{m}x\ne0\right\} $.
Note that $s$ exists, if $\chi(x)\ne1\;\forall x$. Consider arbitrary
$x\in\mathbf{Q}_{m}$ \[
x=m^{-r}\left(b_{0}+mb_{1}+m^{2}b_{2}+...+m^{r-s-1}b_{r-s-1}\right)+m^{-s}\left(a_{0}+ma_{1}+m^{2}a_{2}+...\right)\]
 As $\chi(B_{s})=1$ we get \[
\chi(x)=\left(\chi\left(m^{-r}\right)\right)^{b_{0}}\left(\chi\left(m^{-r+1}\right)\right)^{b_{1}}...\left(\chi\left(m^{-s-1}\right)\right)^{b_{r-s-1}}\]
 After some transformations \[
\chi\left(m^{-r}\right)=\chi\left(m^{-s+s-r}\right)=\left(\chi\left(m^{-s}\right)\right)^{m^{s-r}}=\exp\left(2\pi i\left(m^{-r+s}l\right)\right),\]
 \[
\chi\left(m^{-r+1}\right)=\chi\left(m^{-r}\right)^{m}=\exp\left(2\pi i\left(m^{-r+s+1}l\right)\right),\]
 \[
\cdots\]
 \[
\chi\left(m^{-s-1}\right)=\left(\chi\left(m^{-r}\right)\right)^{m^{r-s-1}}=\exp\left(2\pi i\left(m^{-1}l\right)\right),\]
 where $l=0,1,...,m^{r-s}-1,$ we get \[
\chi(x)=\exp\left(2\pi i\left(\left(lm^{s}\right)\left(m^{-r}b_{0}+m^{-r+1}b_{1}+...+m^{-s-1}b_{r-s-1}\right)\right)\right).\]
 By denoting $lm^{s}\equiv k$ and expanding $k=m^{s}\left(k_{0}+mk_{1}+...+m^{r-s-1}k_{r-s-1}\right)$
we have \[
\chi(x)=\]
 \[
=\exp\left[2\pi i\left(\left(m^{s}\left(k_{0}+mk_{1}+...+m^{r-s-1}k_{r-s-1}\right)\right)\right.\right.\times\]
 \[
\left.\left.\left(m^{-r}\left(b_{0}+mb_{1}+...+m^{r-s-1}b_{r-s-1}\right)\right)\right)\right].\]
 Thus every character for $\forall x\in\mathbf{Q}_{m}$ is defined
by $k\in\mathbf{Q}_{m}$ and has the form \[
\chi(x)=\exp\left(2\pi ikx\right)=\exp\left(2\pi i\left\{ kx\right\} \right).\]

Let $f(x)\in L^{1}(\mathbf{Q}_{m})$, then for any $k\in\mathbf{Q}_{m}$
$f(x)\chi_{m}(kx)\in L^{1}(\mathbf{Q}_{m})$. Let us call $m$-adic
Fourier transform of a function $f(x)$such function $\tilde{f}(k)$
that \[
\tilde{f}(k)=\intop_{\mathbf{Q}_{m}}\chi_{m}(kx)f(x)d_{m}x.\]
 Function $f(x)$ on $\mathbf{Q}_{m}$ is called locally constant
if $f(x)$ is finite for all $x\in\mathbf{Q}_{m}$, and for all $x\in\mathbf{Q}_{m}$
there exists integer $l(x)$ that $f(x+x^{\prime})=f(x)$ for any
$x^{\prime}\in B_{l}(x)$. 
If there exists integer $l$ such that
for all $x\in\mathbf{Q}_{m}$ and for
all $x^{\prime}\in B_{l}$
the equation $f(x+x^{\prime})=f(x)$
holds,
then $f(x)$ is called locally constant on
$\mathbf{Q}_{m}$ with constancy parameter $l$.
Let us implement
the following designations:

$E(\mathbf{Q}_{m})$ is set of functions locally constant on $\mathbf{Q}_{m}$;

$E^{l}(\mathbf{Q}_{m})$ is set of functions locally constant on $\mathbf{Q}_{m}$
with constancy parameter $l$;

$D(\mathbf{Q}_{m})$ is set of functions on $\mathbf{Q}_{m}$with
compact support;

$D_{r}(\mathbf{Q}_{m})$ is set of functions on $\mathbf{Q}_{m}$with
a support in a disk $B_{r}$;

$D_{r}^{l}(\mathbf{Q}_{m})$ is set of functions locally constant
on $\mathbf{Q}_{m}$ with support in disk $B_{r}$ and constancy parameter
$l$.

\textbf{Theorem 4.2.} \textit{If $f(x)\in D_{r}^{l}(\mathbf{Q}_{m})$,
then $\tilde{f}(k)\in D_{-l}^{-r}(\mathbf{Q}_{m})$.}

\textbf{Proof. }Let $k^{\prime}\in B_{-r}$. Then \[
\tilde{f}(k+k^{\prime})=\intop_{\mathbf{Q}_{m}}\chi_{m}(kx+k^{\prime}x)f(x)d_{m}x=\int_{B_{r}}\chi_{m}(kx)\chi_{m}(k^{\prime}x)f(x)d_{m}x\]
 Because in the integrand $|k^{\prime}x|_{m}\le|k^{\prime}|_{m}|x|_{m}\le1$,
then $\chi_{m}(k^{\prime}x)=1$, we have $\tilde{f}(k+k^{\prime})=\int_{B_{r}}\chi_{m}(kx)f(x)d_{m}x=\int_{\mathbf{Q}_{m}}\chi_{m}(kx)f(x)d_{m}x=\tilde{f}(k)$
and $\tilde{f}(k)\in E^{-r}(\mathbf{Q}_{m})$. Let $|k|_{m}>m^{-l}$.
Then after substitution $x\rightarrow y=x-m^{-l}$ we get \[
\tilde{f}(k)=\intop_{\mathbf{Q}_{m}}\chi_{m}(kx)f(x)d_{m}x=\intop_{\mathbf{Q}_{m}}\chi_{m}(ky+km^{-l})f(y+m^{-l})d_{m}y=\]
 \[
=\chi_{m}(km^{-l})\intop_{\mathbf{Q}_{m}}\chi_{m}(ky)f(y)d_{m}y=\chi_{m}(km^{-l})\tilde{f}(k).\]
 Because $|km^{-l}|_{m}=|k|_{m}|m^{-l}|_{m}=|k|_{m}m^{l}>1$, we have
$\left(1-\chi_{m}(km^{-l})\right)\ne0$ hence $\tilde{f}(k)=0$ and
$\tilde{f}(k)\in D_{-l}^{-r}(\mathbf{Q}_{m})$.

\textbf{Theorem 4.3.} \textit{Let a function $f(x)$ be continuous
on $\mathbf{Q}_{m}$ and $f(x)\in L^{1}(\mathbf{Q}_{m})$. Then the
inversion formula holds:} \[
f(x)=\intop_{\mathbf{Q}_{m}}\chi_{m}(-kx)\tilde{f}(k)d_{m}k.\]

\textbf{Proof. }Consider \[
f_{r}(x)\equiv\int_{B_{r}}\chi_{m}(-kx)\left(\intop_{\mathbf{Q}_{m}}\chi_{m}(ky)f(y)d_{m}y\right)d_{m}k.\]
 Let us show that $\lim\limits _{r\to\infty}f_{r}(x)=f(x)$. By uniform
convergence of the internal integral we have \[
f_{r}(x)=\intop_{\mathbf{Q}_{m}}\int_{B_{r}}\chi_{m}\left(k(y-x)\right)d_{m}kf(y)d_{m}y=\]
 \[
=m^{r}\intop_{\mathbf{Q}_{m}}\Omega(m^{r}|y-x|_{m})f(y)d_{m}y,\]
 where $\Omega(\lambda)=\left\{ \begin{array}{l}
1,\:\lambda\le1,\\
0,\:\lambda>1.\end{array}\right.$ Taking into account $m^{-r}\int_{\mathbf{Q}_{m}}\Omega(m^{r}|y-x|_{m})d_{m}y=1$,
we have\textit{\ \[
f_{r}(x)-f(x)=m^{r}\intop_{\mathbf{Q}_{m}}\Omega(m^{r}|y-x|_{m})\left(f(y)-f(x)\right)d_{m}y=\]
 \[
=m^{r}\intop_{|y-x|_{m}\le m^{-r}}\left(f(y)-f(x)\right)d_{m}y.\]
 }Due to continuity of $f(x)$ one gets $\lim\limits _{r\to\infty}\left(f_{r}(x)-f(x)\right)=0$.

\textbf{Theorem 4.4.} \textit{If the functions $f(x)$ and $g(x)$
are continuous and bounded on $\mathbf{Q}_{m}$ and $f(x)\in L^{1}(\mathbf{Q}_{m})$,
$g(x)\in L^{1}(\mathbf{Q}_{m})$ then } \[
\intop_{\mathbf{Q}_{m}}\chi_{m}(kx)\left(f*g\right)(x)d_{m}x=\tilde{f}(k)\tilde{g}(k).\]

\textbf{Proof. }Let us apply Theorem 3.1 replacing $f(x)$ and $g(x)$
with $\chi_{m}(kx)f(x)$ and $\chi_{m}(kx)g(x)$: \[
\intop_{\mathbf{Q}_{m}}\left[\intop_{\mathbf{Q}_{m}}\left(\chi_{m}(ky)f(y)\right)\left(\chi_{m}(k(x-y))g(x-y)\right)d_{m}y\right]d_{m}x=\]
 \[
=\intop_{\mathbf{Q}_{m}}\chi_{m}(kx)f(x)d_{m}x\intop_{\mathbf{Q}_{m}}\chi_{m}(kx)g(x)d_{m}x.\]
 Then we obtain \[
\intop_{\mathbf{Q}_{m}}\chi_{m}(kx)\left(\int_{\mathbf{Q}_{m}}f(y)g(x-y)d_{m}y\right)d_{m}x=\intop_{\mathbf{Q}_{m}}\chi_{m}(kx)f(x)d_{m}x\intop_{\mathbf{Q}_{m}}\chi_{m}(kx)g(x)d_{m}x.\]

\section{Levi processes with a value in $\mathbf{Q}_{\boldsymbol{m}}$}

According to Kolmogorov's axiomatic \cite{key-17}, by probability
space is understood a~trip\-let of the elements \textbf{$\left\{ \Omega,\Sigma,P\right\} $},
where \textbf{$\left\{ \Omega,\Sigma\right\} $ } is measurable space,
$P$- non-negative countable additive function on $\sigma$-algebra
$\Sigma$ satisfying the condition $P(\Omega)=1$. Let~$\left\{ \Omega,\Sigma\right\} $
be measurable space and $\left\{ \mathbf{Q}_{m},\mathrm{B}\right\} $
is the pair composed of a ring of $m$-adic numbers $\mathbf{Q}_{m}$
and $\sigma$-algebra $\mathrm{B}$ of all measurable subsets of $\mathbf{Q}_{m}$.
Mapping $X:\Omega\to\mathbf{Q}_{m}$ is called $\Sigma|\mathrm{B}$
-- measurable (or $X\in\Sigma|\mathrm{B}$) if $X^{-1}(\mathrm{B})\subset\Sigma$.
$m$-Adic random variable $X=X(\omega)$ is $\Sigma|{\ \mathrm{B}}$
-- measurable mapping $X:\Omega\to\mathbf{Q}_{m}$. Function $X(\omega)$
generates the probability measure $P_{X}(B)$ on $B\in\mathrm{B}$
by relation $P_{X}(B)=P\left(X^{-1}(B)\right)$. The function $P_{X}(B)$
is called distribution function of $m$-adic random variable $X$.
$X(\omega)$ is continuous, if there exists integrable on any $B\in\mathrm{B}$
function $f(x)$ such that $P_{X}(B)=\int_{B}f(x)d_{m}x$. We shall
call function $f(x)$ density of distribution function of $m$-adic
random variable $X$. In the simplest case it is possible to choose
$\Omega\equiv\mathbf{Q}_{m}$ as the probability space $\Sigma\equiv\mathrm{B}$,
$P(B)\equiv\int_{B}f(x)d_{m}x$ for any$B\in\mathrm{B}$ and some
function $f(x)$ such that $f(x)\in L_{\mathrm{loc}}^{\mathrm{1}}(\mathbf{Q}_{m})$,
$f(x)\in L^{1}(\mathbf{Q}_{m})$ and $\int_{\mathbf{Q}_{m}}f(x)d_{m}x=1$.
In such case $X(\omega)=\omega$, $\omega\in\mathbf{Q}_{m}$, and
random variable $X$ is uniquely defined by the density of distribution
function $f(x)$.

Random variable $X$ with the values in $\mathbf{Q}_{m}$ is called
infinitely divisible if for any $n\ge1$ it is possible to find independent
identically distributed random variables $X_{n1},...,X_{nn}$ with the
values in $\mathbf{Q}_{m}$ such that $X=X_{n1}+...+X_{nn}$.

Characteristic function $\mu(t)$ $m$-adic random variable is defined
as \[
\hat{f}(k)=\intop_{\mathbf{Q}_{m}}f(x)\chi(kx)d_{m}x.\]

\textbf{Theorem 5.1.} \textit{Characteristic function of any infinitely
divisible }$m$\textit{-adic distribution can be represented in the
form of (analogous to Levi--Hinchin formula) } \begin{equation}
\hat{f}(k)=\chi(kx_{0})\Omega\left(|k|_{m}p^{-r}\right)\exp\left(\intop_{\mathbf{Q}_{m}\backslash\{0\}}\left(\chi(kx)-1\right)W(x)d_{m}x\right),\label{L-H}\end{equation}
 \textit{where }$x_{0}\in\mathbf{Q}_{m}$\textit{, }$r$\textit{ is
integer, }$W(x)d_{m}x$\textit{ \ is the measure on }$\mathbf{Q}_{m}$\textit{,
finite on }$\mathbf{Q}_{m}\backslash B_{s}$\textit{ for any }$s$\textit{,
for which }$\intop_{\mathbf{Q}_{m}\backslash\{0\}}\left(1-Re\chi(kx)\right)W(x)d_{m}x<\infty$\textit{
and}\textit{\emph{ }}\emph{here} $\int_{\mathbf{Q}_{m}\backslash\{0\}}\cdots\, d_{m}x=\lim\limits _{r\rightarrow-\infty}\int_{\mathbf{Q}_{m}\backslash B_{r}}\cdots\, d_{m}x$\textit{
is the principal value of the integral.}

\textit{\emph{The proof of the Theorem 5.1 coincides with the proof
of analogous theorem for totally disconnected locally compact Abel
groups (e.g. refer to \cite{key-18}) also for the field }}\emph{$Q_{p}$}\textit{\emph{
\cite{key-19}.}}

Let us notice that the homogeneous distributions and rotation-invariant
temporally and spatially homogeneous processes on $\mathbf{Q}_{p}^{n}$ with
independent increments were considered in resent paper~\cite{key-add}.

By $m$-adic number random process we shall call mapping $X=X(t,\omega)$of
$\mathrm{\mathbf{R}}_{+}\otimes\Omega$ on $\mathbf{Q}_{m}$ which
is $\Sigma|\mathrm{B}$-measurable for any $t\in\mathbf{R}_{+}$.

By Levi process $X(t)$ from $\mathbf{Q}_{m}$ we shall call stochastically
continuous Markov process with stationary independent increments,
beginning from zero and having right continuous trajectories with
left side limits, i.e. process for which the following conditions
are hold:
\begin{enumerate}
\item $X(0)=0$ almost everywhere.
\item For any $0\le t_{1}<...<t_{n}$ increments $X(t_{1})$, $X(t_{2})-X(t_{1})$,
$\ldots$ , $X(t_{n})-X(t_{n-1})$ are independent.
\item For any $t$ and $s$$X(t+s)-X(s)=X(t)-X(0)$.
\item For any $t\ge0$ and $r$ $\lim\limits _{s\to t}P\left(|X(s)-X(t)|_{m}>p^{r}\right)=0$.
\item Trajectories $X(t)$ for all $t>0$ almost everywhere right continuous
and possess left limits.
\end{enumerate}
From the definition of Levi process it follows that $X(t)$ is infinitely
divisible. Characteristic function of Levi process is defined as \[
\hat{f}(k,t)=\mathrm{M}\left[\chi\left(kX(t)\right)\right]\]
 and due to uniformity and independence of increments has the form
\[
\hat{f}(k,t)=\exp\left(t\psi(k)\right),\]
 where the form of the function $\psi(k)$ follows from (\ref{L-H})
\[
\psi(k)=\intop_{\mathbf{Q}_{m}\backslash\{0\}}\left(\chi(kx)-1\right)W(x)d_{m}x\]
 It is not difficult to see that $\hat{f}(k,t)=\exp\left(t\intop_{\mathbf{Q}_{m}\backslash\{0\}}\left(\chi(kx)-1\right)W(x)d_{m}x\right)$
satisfies the equation \begin{equation}
\frac{d}{dt}\hat{f}(k,t)=\hat{f}(k,t)\intop_{\mathbf{Q}_{m}\backslash\{0\}}\left(\chi(kx)-1\right)W(x)d_{m}x.\label{K-F_F}\end{equation}

From (\ref{K-F_F}) the equations for the density of distribution
function for the Levi process follows \begin{equation}
\frac{d}{dt}f(x,t)=\intop_{\mathbf{Q}_{m}\backslash\{0\}}W(y-x)\left(f(y,t)-f(x,t)\right)d_{m}y\label{K-F}\end{equation}

Thus the following theorem takes place.

\textbf{Theorem 5.2}\textbf{\textit{.}}\textit{\ Distribution function
}$f(x,t)$\textit{\ for all Levi process with the values in }$\mathrm{\mathbf{Q}}_{m}${
}\textit{satisfies eq. (\ref{K-F}) where }$W(x)d_{m}x$\textit{
is the measure on }$\mathrm{\mathbf{Q}}_{m}$\textit{ finite on }$\mathrm{\mathbf{Q}}_{m}\backslash B_{s}$\textit{
for all }$s$\textit{, for which }$\intop_{\mathbf{Q}_{m}\backslash\{0\}}\left(1-Re\chi(kx)\right)W(x)d_{m}x<\infty$\textit{.}

Note that a particular case of Levi process in $\mathbf{Q}_{m}$ is
the one described by Vladimirov equation for which ${\displaystyle W(x)=-\frac{1}{\Gamma_{m}(-\alpha)}\frac{1}{|x|^{\alpha}}}$,
$\Gamma_{m}(-\alpha)=\frac{1-m^{-\alpha-1}}{1-m^{\alpha}}$, $\alpha>0$:
\begin{equation}
\frac{df(x,t)}{dt}=-\frac{1}{\Gamma_{m}(-\alpha)}\intop_{\mathbf{Q}_{m}\backslash\{0\}}\frac{f(y,t)-f(x,t)}{|y-x|_{m}^{\alpha+1}}d_{m}y.\label{Vlad}\end{equation}

\section{ $\mathbf{Q}_{\boldsymbol{m}}$-valued uncoupled continuous time
random walk}

Let $\left(\xi_{1},T_{1}\right),\,\left(\xi_{1},T_{1}\right),\ldots$
be a sequence of independent identically distributed pairs of random
variables $\xi_{i}\in\mathbf{Q}_{m}$ and $T_{i}\in\mathrm{R}_{+}$.
Let the distribution of each pair is given by probability measure
$P(d_{m}x,dt)$ on $\mathbf{Q}_{m}\times\mathrm{R}_{+}$. We shall
suppose that $\xi_{i}$ and $T_{i}$ are independent and hence $P(d_{m}x,dt)=\varphi(x)d_{m}x\psi(t)dt$
where $\varphi(x)$ and $\psi(t)$ are probability densities of $\xi_{i}$
and $T_{i}$ respectively. Let us define random process $N(t)$ as
\begin{equation}
N(t)=\max\left\{ n:\:\sum_{i=1}^{n}T_{i}\le t\right\} .\label{N(t)}\end{equation}
 We shall call the process \begin{equation}
X(t)=\sum_{i=1}^{N(t)}\xi_{i}\label{X(t)}\end{equation}
 as $m$-adic uncoupled continuous time random walk ($m$-adic CTRW).

Let us denote $\sum_{i=1}^{n}T_{i}\equiv t_{n}$. Distribution $p(t,n)$
for $N(t)$ is the measure function $p(t,n)=P(A)$ where $\Sigma\supset A=\left\{ \left\{ \omega\right\} :\;\forall\omega\in A\,,\mathrm{\;\;}t_{n}(\omega)<t\le t_{n+1}(\omega)\right\} $
or, equivalently as mathematic expectation $p(t,n)=\mathrm{M}\left[I\left(t_{n}<t\le t_{n+1}\right)\right]$
where \[
I\left(t_{n}<t\le t_{n+1}\right)=\left\{ \begin{array}{l}
1,\; t_{n}<t\le t_{n+1},\\
\\0,\; t_{n}\ge t\:\textrm{or}\: t>t_{n+1}.\end{array}\right.\]
 Let $\hat{p}(s,n)=L\left[p(t,n)\right](s)$ be Laplace image of $p(t,n)$.
Then \[
\hat{p}(s,n)=\mathrm{M}\left[\intop_{0}^{\infty}dte^{-st}I\left(t_{n}<t<t_{n+1}\right)\right]=\mathrm{M}\left[\frac{e^{-st_{n+1}}-e^{-st_{n}}}{s}\right].\]
 Because $t_{n}=\sum_{i=1}^{n}T_{i}$ is the sum of independent random
variables, we have \[
\mathrm{M}\left[e^{-st_{n}}\right]=\mathrm{M}\left[\exp\left(-s\sum_{i=1}^{n}T_{i}\right)\right]=\hat{\psi}^{n}(s)\]
 where $\hat{\psi}(s)=L\left[\psi(t)\right](s)$. Thus \begin{equation}
\hat{p}(s,n)=\mathrm{M}\left[\frac{e^{-st_{n+1}}-e^{-st_{n}}}{s}\right]=\hat{\psi}^{n}(s)\frac{1-\hat{\psi}(s)}{s}\label{p(s,n)}\end{equation}
 Distribution density $f(x,t)$ of a process $X(t)$ is Fourier image
of characteristic function of the process $X(t)$ \[
f(x,t)=F^{-1}\left[\tilde{f}(k,t)\right].\]
 Using (\ref{X(t)}), we obtain \[
\tilde{f}(k,t)=\mathrm{M}\left[\chi\left(k\sum_{i=1}^{N(t)}\xi_{i}\right)\right]=\sum_{i=0}^{\infty}\tilde{\varphi}^{n}(k)p(n,t),\]
 where $\tilde{\varphi}(k)=\mathrm{M}\left[\chi\left(\xi_{i}k\right)\right]$
is characteristic function of independent identically distributed
variables $\xi_{i}$. Using (\ref{p(s,n)}) we have \begin{equation}
\hat{\tilde{f}}(k,s)=\frac{1-\hat{\psi}(s)}{s}\frac{1}{1-\hat{\psi}(s)\tilde{\varphi}(k)}.\label{M-W}\end{equation}
 Formula (\ref{M-W}) is Monthroll--Weiss equation \cite{key-20}
for $m$-adic CTRW with independent jumps. After simple transformations,
the equation (\ref{M-W}) can be represented in an alternative form
\begin{equation}
\hat{\Phi}(s)\left[s\hat{\tilde{f}}(k,s)-1\right]=\left[\tilde{\varphi}(k)-1\right]\hat{\tilde{f}}(k,s),\label{AF_M-W}\end{equation}
 where \[
\hat{\Phi}(s)=\frac{1-\hat{\psi}(s)}{s\hat{\psi}(s)}.\]
 Inverting Laplace--Fourier transforms, it is possible to rewrite
eq. (\ref{AF_M-W}) in the following forms:
\begin{enumerate}
\item Integral equation\[
f(x,t)=\Psi(t)\delta(x)+\intop_{0}^{t}\psi(t-t^{\prime})dt\intop_{\mathbf{Q}_{m}}\varphi(k-k^{\prime})f(x,t)d_{p}k,\:\Psi(t)=L^{-1}\left[\frac{1-\hat{\psi}(s)}{s}\right].\]

\item Generalized Kolmogorov--Feller equation\[
\intop_{0}^{t}\Phi(t-t^{\prime})\frac{d}{dt^{\prime}}f(x,t)dt^{\prime}=\intop_{\mathbf{Q}_{m}}d_{p}x^{\prime}\varphi(x-x^{\prime})\left(f(x^{\prime},t)-f(x,t)\right),\:\Phi(t)=L^{-1}\left[\hat{\Phi}(s)\right].\]

\end{enumerate}
For exponential distribution of time intervals between the jumps \[
\psi(t)=e^{-\lambda t},\:\psi(s)=\frac{\lambda}{s+\lambda},\:\Phi(t)=\frac{1}{\lambda}\delta(t)\]
 the process (\ref{X(t)}) is Levi process with finite measure function
$W(x)d_{m}x\equiv\lambda\varphi(x)d_{m}x$ on the whole $\mathbf{Q}_{m}$.

\section{Diffusive limit of $\mathbf{Q}_{\boldsymbol{m}}$-valued CTRW and
relationship with Vladimirov equation }

Let us multiply all waiting times $T_{i}$ by the factor $r$, and
$\xi_{i}\in\mathbf{Q}_{m}$ -- by the factor $h\in\mathbf{Q}_{m}$.
Then the distribution densities $\psi(t)$ and $\varphi(x)$ are transformed
into \[
\psi_{r}(t)=r^{-1}\psi\left(\dfrac{t}{r}\right),\:\varphi_{h}(x)=|h|_{p}^{-1}\varphi\left(\dfrac{x}{h}\right).\]
 For Laplace and Fourier transforms of $\psi(t)$ and $\varphi(x)$
we have \[
\hat{\psi}_{r}(s)=\hat{\psi}(rs),\:\tilde{\varphi}_{h}(k)=\tilde{\varphi}(hk).\]
 Then Monthroll--Weiss equation (\ref{M-W}) will have the form \[
\hat{\tilde{f}}_{r,h}(k,s)=\frac{1-\hat{\psi}_{r}(s)}{s}\frac{1}{1-\hat{\psi}_{r}(s)\tilde{\varphi}_{h}(k)}=\frac{1-\hat{\psi}(rs)}{s}\frac{1}{1-\hat{\psi}(rs)\tilde{\varphi}(hk)}\]
 We propose that $\hat{\psi}(s)$ and $\tilde{\varphi}(k)$ have the
following asymptotic behavior at $s\to0$, $|k|_{m}\to0$: \[
\hat{\psi}(s)=1-s^{\beta}+o(s^{\beta}),\:0<\beta\le1,\tilde{\:\varphi}(k)=1-|k|_{m}^{\alpha}+o\left(|k|_{m}^{\alpha}\right),\:\alpha>0\]
 Then we define diffusive limit of $m$-adic CTRW (similarly to definition
of diffusive limit of real value CTRW \cite{key-21,key-22}) as \[
r\to0,\:|h|_{m}\to0\:\textrm{ ~ on condition that }\: r^{\beta}=|h|_{m}^{\alpha}.\]
 It is not difficult to see that \[
\hat{\tilde{f}}_{r,h}(k,s)\to\hat{\tilde{u}}(k,s)\equiv\frac{1}{s^{\beta}+|k|_{m}^{\alpha}}.\]
 In this passage to the limit, $\hat{\tilde{f}}_{r,h}(k,s)$and $\hat{\tilde{u}}(k,s)$
are asymptotically equivalent in the Laplace--Fourier domain. Then,
the asymptotic equivalence in the \textbf{$\mathbf{Q}_{m}$-}space
- real time domain between the $f_{r,h}(x,t)$ and $u(x,t)\equiv F^{-1}L^{-1}\left[\hat{\tilde{u}}(k,s)\right]$
is ensured by the continuity theorem for sequences of characteristic
functions, after the application of the analogous theorem for sequences
of Laplace transforms \cite{key-23}. Exactly in the same way as in
the case of diffusive limit of real valued CTRW, it is possible to
show that (ref. 
\cite{key-21,key-22}) that \[
f_{r,h}(x,t)\to u(x,t).\]
 Precise evaluations show that \[
u(x,t)=F^{-1}L^{-1}\left[\hat{\tilde{u}}(k,s)\right]=\intop_{\mathbf{Q}_{m}}\chi\left(-kx\right)E_{\beta}(-|k|_{m}^{\alpha}t^{\beta})d_{m}k=\]
 \begin{equation}
=(1-m^{-1})|x|_{m}^{-1}\sum_{i=0}^{\infty}m^{-i}E_{\beta}(-m^{-i\alpha}|x|_{m}^{-\alpha}t^{\beta})-|x|_{m}^{-1}E_{\beta}(-m^{\alpha}|x|_{m}^{-\alpha}t^{\beta}),\label{u(x,t)}\end{equation}
 where \[
E_{\beta}(z)=\sum_{n=0}^{\infty}\frac{z^{n}}{\Gamma(\beta n+1)}\]
 is Mittag-Leffler function. Function (\ref{u(x,t)}) satisfies the
equation \begin{equation}
\frac{\partial^{\beta}}{\partial t^{\beta}}u(x,t)=-\frac{1}{\Gamma_{m}(-\alpha)}\intop_{\mathbf{Q}_{m}}\frac{u(y,t)-u(x,t)}{|x-y|_{m}^{\alpha+1}}d_{m}y,\label{FT_EQ}\end{equation}
 where $\dfrac{\partial^{\beta}}{\partial t^{\beta}}$ is the fraction
Caputo derivative \cite{key-24} \[
\frac{\partial^{\beta}}{\partial t^{\beta}}f(t)\equiv\frac{1}{\Gamma(1-\beta)}\left(\frac{\partial}{\partial t}\intop_{0}^{t}\frac{f(\tau)}{(t-\tau)^{\beta}}d\tau-t^{-\beta}f(0)\right).\]
 We shall call equation (\ref{FT_EQ}) the equation of fractional-time
$m$-adic random walk. For $\beta=1$, eq. (\ref{FT_EQ}) becomes
the $m$-adic Vladimirov equation (\ref{Vlad}).

\section{The equation of fraction-time $\boldsymbol{m}$-adic random walk
and asymptotic behavior of the probability measure of initial distribution
support}

We consider the equation of fractional-time $m$-adic random walk
(\ref{FT_EQ}) for $0<\beta\le1,\:\alpha>0$ with the initial
condition $u(x,0)=\Omega(|x|_{m}\le1)$. Solution of eq. (\ref{FT_EQ})
is easy to find using Laplace--Fourier transformations: \[
u(x,t)=\Omega\left(|x|_{m}\le1\right)\left(1-m^{-1}\right)\sum_{i=0}^{\infty}m^{-r}E_{\beta}(-m^{-\alpha i}t^{\beta})+\]
 \[
+\Omega\left(|x|_{m}>1\right)\left(\left(1-m^{-1}\right)|x|_{m}^{-1}\sum_{i=0}^{\infty}m^{-i}E_{\beta}(-m^{-\alpha i}|x|_{m}^{-\alpha}t^{\beta})-|x|_{m}^{-1}E_{\beta}(-m^{\alpha}|x|_{m}^{-\alpha}t^{\beta})\right).\]
 Probability measure of the support of initial distribution is defined
as \begin{equation}
S(t)=\intop_{Z_{m}}u(x,t)d_{m}x=\left(1-m^{-1}\right)\sum_{i=0}^{\infty}m^{-i}E_{\beta}(-m^{-\alpha i}t^{\beta}).\label{S(t)}\end{equation}
 Let us find the estimation of (\ref{S(t)}) at $t\to\infty$. Note
that the function $m^{-x}$ decreases with the increase of $x$, while
$E_{\beta}(-m^{-\alpha x}t^{\beta})$ increases. Consequently on the
segment $i\le x\le i+1$ we have in equation \begin{equation}
m^{-x}E_{\beta}(-m^{-i\alpha(x-1)}t^{\beta})\le m^{-i}E_{\beta}(-m^{-i\alpha}t^{\beta})\le m^{-(x-1)}E_{\beta}(-m^{-\alpha x}t^{\beta}).\label{EQ1}\end{equation}
 Integrating (\ref{EQ1} over $x$ from $i$ to $i+1$ and then summing
by $i$ from $0$ to $\infty$, we obtain \[
\frac{1}{\alpha p\ln p}t^{-\frac{\beta}{\alpha}}\mathrm{\;}\Theta_{\beta}(\alpha,t)\le S(t)\le\frac{p}{\alpha\ln p}t^{-\frac{\beta}{\alpha}}\mathrm{\;}\Theta_{\beta}(\alpha,t),\]
 where \[
\Theta_{\beta}(\alpha,t)\equiv\intop_{0}^{t^{\beta}}E_{\beta}(-y)y^{\frac{1}{\alpha}-1}dy,\:\Theta_{\beta}(\alpha)\equiv\intop_{0}^{\infty}E_{\beta}(-y)y^{\frac{1}{\alpha}-1}dy.\]
 Because at $t\to+\infty$ and $\lambda>0$ the following asymptotic formula holds \[ 
E_{\beta}(-\lambda t^{\beta})=\frac{\sin(\beta\pi)}{\lambda\pi}\frac{\Gamma(\beta)}{t^{\beta}}\left(1+o(1)\right),\]
 the integral in $\Theta_{\beta}(\alpha)$ converges at $\alpha>1$
and we have \[
\frac{1}{\alpha p\ln p}t^{-\frac{\beta}{\alpha}}\Theta_{\beta}(\alpha)\left(1+o(1)\right)\le S(t)\le\frac{p}{\alpha\ln p}t^{-\frac{\beta}{\alpha}}\Theta_{\beta}(\alpha)\left(1+o(1)\right).\]
 At $\alpha\le1$ when $y\to+\infty$ we have \[
E_{\beta}(-y)y^{\frac{1}{\alpha}-1}=\frac{\sin(\beta\pi)\Gamma(\beta)}{\pi}y^{\frac{1}{\alpha}-2}\left(1+o(1)\right),\]
 therefore \[
\intop_{0}^{t^{\beta}}E_{\beta}(-y)y^{\frac{1}{\alpha}-1}dy=\left\{ \begin{array}{l}
\dfrac{\sin(\beta\pi)\Gamma(\beta)}{\pi}t^{\beta\left(\frac{1}{\alpha}-1\right)}\left(1+o(1)\right),\:\alpha<1,\\
\\\dfrac{\sin(\beta\pi)\Gamma(\beta)}{\pi}\ln t\left(1+o(1)\right),\:\alpha=1.\end{array}\right.\]
 Finally we obtain the following asymptotic estimate of $S(t)$ \[
\frac{1}{\alpha p\ln p}R(t)\le S(t)\le\frac{p}{\alpha\ln p}R(t),\]
 where at $t\to+\infty$ \[
R(t)=\left\{ \begin{array}{l}
t^{-\frac{\beta}{\alpha}}\Theta_{\beta}(\alpha)\left(1+o(1)\right),\:\alpha>1,\\
\\\dfrac{\sin(\beta\pi)\Gamma(\beta)}{\pi}t^{-\beta}\left(1+o(1)\right),\:\alpha<1,\\
\\\dfrac{\sin(\beta\pi)\Gamma(\beta)}{\pi}\ln t\left(1+o(1)\right),\:\alpha=1.\end{array}\right.\]

\section{Summary}

In this paper we developed a number of theorems from the $p$-adic
\cite{key-2,key-13} into $m$-adic 
analysis of complex valued function. 
This allowed us to consider some type of $m$-adic-valued stochastic
processes. In particular, 
the distribution function of any $m$-adic-valued Levy process satisfies
the pseu\-do\-dif\-ferential equation 
of generalized ultrametric random walk. The class of $m$-adic CTRW
random processe was considered 
and 
its diffusion limit was obtained. 
It is well known that diffusion limit of real valued CTRW (fractional-time
in general) is the diffusion process. However the trajectories of
ultrametric processes are not continuous and any of such process is
not diffusion in the traditional definition, because the Lindeberg
condition is not hold. But we can define $m$-adic process, being
the diffusion limit of $m$-adic CTRW, as diffusion indeed. Nevertheless
we do not follow this terminology. We call the diffusion limit of
$m$-adic CTRW as fractional-time $m$-adic random walk. The distribution
function of such process satisfies the Vladimirov equations with fractional
time derivative. We also obtained the fractional-time $m$-adic random
walk distributive function in explicit form and check the asymptotic
behavior of the probability measure of the initial distribution support
at large times.

It is interesting to consider the next generalization of the $m$-adic
analysis to polyadic. As it is known \cite{key-25} for any integers 
sequence $\dots,\; a_{-2},$ ${\rm \;\;}a_{-1},$ ${\rm \;\;}a_{0},{\rm \;\;}a_{1},{\rm \;\;}a_{2},\;\dots$, ($a_{i} > 1)$
there exist the set of polyadic ($a$-adic) numbers $\mathbf{Q}_{\{a\}}$ that
is the set of formal series $x=\dots+\dfrac{1}{a_{-3}a_{-2}a_{-1}}x_{-3}+\dfrac{1}{a_{-2}a_{-1}}x_{-2}+\dfrac{1}{a_{-1}}x_{-1}+x_{0}+a_{0}x_{1}+a_{1}a_{0}x_{1}+\dots,$
with $x_{i}=0,1,\dots,a_{i}-1$. The set $\mathbf{Q}_{\{a\}}$ is
local compact abelian commutative group and has Haar measure and character
group, that admits exact realization of measure, Fourier analysis
and detailed consideration of polyadic stochastic processes.

\vskip 0.5cm
\noindent
{\large \bf Acknowledgements}
\vskip 0.2cm
\noindent
A.Z. is grateful to Igor Volovich, Sergey Kozyrev, Vladik Avetisov, Albert Bikulov, Luigi Accardi and
Branko Dragovich for useful discussions. Work was partially supported by grants ADTP 3341, 10854
and RFBR 09-01-12161-ofi-m.

\end{document}